\documentclass[prl,superscriptaddress,aps, bibliography, twocolumn, reprint, showpacs, footnoteinbib]{revtex4-1}
\usepackage{graphicx}
\usepackage{amssymb}   % for math
\usepackage{amsfonts}
\usepackage{amsmath}
\usepackage{dsfont}
\usepackage{natbib}
\usepackage{dcolumn}   % needed for some tables
\usepackage{bm}        % for math
\usepackage[mathscr]{eucal}
\usepackage[dvipsnames]{xcolor}
\usepackage[colorlinks, linkcolor=OrangeRed,citecolor=RoyalBlue,urlcolor=NavyBlue]{hyperref}
\usepackage[all]{hypcap} % let hyperlinks correctly point to figures rather than their captions;
\usepackage{mathtools}
\usepackage{wasysym}

\definecolor{myblue}{rgb}{0,0,0.75}

\usepackage{graphicx}
\usepackage{amsmath,amssymb,bm}
\usepackage{enumerate}
\usepackage{braket}        % Dirac notation

\begin{document}

\title{Solving efficiently the dynamics of many-body localized systems at strong disorder}
\author{Giuseppe De Tomasi}
\affiliation{Technische Universit\"at M\"unchen, 85747 Garching, Germany}
\affiliation{Max-Planck-Institut f\"ur Physik Komplexer Systeme, N\"othnitzer Stra{\ss}e 38,  01187-Dresden, Germany}
\author{Frank Pollmann}
\affiliation{Technische Universit\"at M\"unchen, 85747 Garching, Germany}
\affiliation{Munich Center for Quantum Science and Technology (MCQST), Schellingstr. 4, D-80799 M\"unchen, Germany}
\author{Markus Heyl}
\affiliation{Max-Planck-Institut f\"ur Physik Komplexer Systeme, N\"othnitzer Stra{\ss}e 38,  01187-Dresden, Germany}
\begin{abstract}
We introduce a method to efficiently study the dynamical properties of many-body localized systems in the regime of strong disorder and weak interactions.
Our method reproduces qualitatively and quantitatively the time evolution with a polynomial effort in system size and independent of the desired time scales.
We use our method to study quantum information propagation, correlation functions, and temporal fluctuations in one- and two-dimensional MBL systems.
Moreover, we outline strategies for a further systematic improvement of the accuracy and we point out relations of our method to recent attempts to simulate the time dynamics of quantum many-body systems in classical or artificial neural networks.
\end{abstract}
\maketitle

{\it Introduction}---
Experiments in quantum simulators, such as ultra-cold atoms in optical lattices and trapped ions, have nowadays achieved access to the dynamical properties of closed 
quantum many-body systems far from equilibrium~\cite{Zwerger08,2012Blatt,2012Bloch,2014Georgescu}. 
Therefore, it has become possible to experimentally study intrinsically dynamical phenomena that are challenging to realize and probe on other platforms. 
One prominent example constitutes the many-body localized phase in systems with strong disorder, whose  signatures have been observed in a series of recent experiments~\cite{schreiber15,Choi16,Smith2016,Bordia16,Lus17}. 
Many-body localization (MBL) describes a nonergodic phase of matter, 
in which particles are localized due to the presence of a strong disorder potential~\cite{Basko06, doi:10.1146/annurev-conmatphys-031214-014726, 2018arXiv180411065A,ALET2018}, 
extending the phenomenon of Anderson localization~\cite{Anderson58} to the interacting case. 
Importantly, the presence of interactions makes the dynamical properties much richer~\cite{Bar12, Proz08, Pep13, Ze18, Canovi11, Ser14, Raj16}.
In particular, interactions give rise to an additional dephasing mechanism, allowing entanglement and quantum information propagation  
even though particle and energy transport is absent~\cite{Bar12, Proz08, Pep13, Ze18,PhysRevB.96.174201}. 
Describing, however, quantitatively this interaction-induced propagation for large systems beyond exact numerical methods has remained as one of the main challenges. 

In this work, we introduce an efficient numerical method to compute the dynamics of weakly-interacting fermions in a fully localized MBL phase.
The method is controlled by the interaction strength and we find that the error remains bounded in time over many temporal decades up to the asymptotic long-time dynamics of quantum information transport in MBL systems, which occurs on time scales exponentially in system size.
The computational resources for computing local observables and correlation functions in our approach scale only polynomially in system size and are even independent of the targeted time in the dynamics.
We utilize the method to study the dynamics of interacting fermions not only in one dimension ($1D$) but also in two dimensions ($2D$) for up to $200$ lattice sites.
After benchmarking our approach by comparing the characteristic entanglement entropy growth with exact diagonalization, we study the 
quantum information transport on the basis of the quantum Fisher information~\cite{PhysRevLett.72.3439,Hauke2016,Smith2016, Petz, 2014Strobel, 2012Toth, 2012Hyllus}, the logarithmic light-cone in correlation functions ~\cite{PhysRevB.96.174201,PhysRevB.97.060201, 2018arXiv180509819L, Cheneau2012, PhysRevA.89.031602, L_uchli_2008, PhysRevB.97.205103, PhysRevLett.113.187203, PhysRevB.96.064303}, and temporal fluctuations of observables, both for $1D$ and $2D$.
Finally, we point out a connection between our approach and recent ideas to encode quantum states into classical and artificial neural networks.
%%%%%%%%%%%%%%%%%%%%%%%%%%%%%%%%%%%%%%%%%%%%

{\it Models $\&$ Methods}---
%The main idea of this work is to construct a 
%simplified model which in the strong localized limit reproduces qualitatively and quantitatively the dynamics 
%of an MBL system and which can be efficiently simulated in a polynomial time in system size. 
At sufficiently strong disorder the MBL eigenstates are expected to be adiabatically connected to the non-interacting ones~\cite{Imbrie2016,Bala13}.
In such a case the system is fully described by an extensive number of quasi-local integral of motions $\{\hat{I}_l\}$~\cite{Chandran15,Ros15,Huse14, Aba13,PhysRevB.97.094206, doi:10.1002/andp.201600322,PhysRevB.97.060201, PhysRevLett.116.160401, Rade16}, which emphasize
an emerging weak form of integrability~\cite{Huse14,Imbrie2016}. In this case the Hamiltonian of the system exhibits a representation of the following form:
\begin{equation}
 \hat{H} = \sum_l J_l^{(1)} \hat{I}_l + \sum_{l,m} J_{l,m}^{(2)} \hat I_l \hat I_m + \dots \, ,
 \label{eq:lbit}
\end{equation} 
where $l$ enumerates the sites of the underlying lattice.
For the considered weakly interacting case, higher-order couplings between the integrals of motion $\hat I_l$ become exponentially suppressed in the interaction strength,  so that we can terminate the expansion as done in Eq.~(\ref{eq:lbit}).
Moreover, it is expected that $J_{l,m}^{(2)}\sim e^{-d(l,m)/\xi}$ with $d(l,m)$ the spatial distance of the two involved lattice sites $l$ and $m$ and $\xi$ denoting the localization length.
While it is expected that this so-called $l$-bit representation exists, it has remained as a central challenge (i) to construct explicitly the integrals of motion $\{\hat I_l\}$ and (ii) to make use of the $l$-bit Hamiltonian to compute its dynamics.

In this work, we show that in the limit of weakly interacting fermions at strong disorder both of these challenges can be efficiently solved.
In this limit we can decompose the Hamiltonian $\hat{H}=\hat{H}_0+\hat{\mathcal{V}}$ with $\hat{H}_0$ a non-interacting Anderson-localized system 
and $\hat{\mathcal{V}}$ the interaction part, whose strength we denote by $V$.
We take as the $\hat I_l$'s the integrals of motion of $\hat H_0 = \sum_l \epsilon_l \hat I_l$
with $\hat I_l = \hat \eta_l^\dag  \hat \eta_l$ and $\hat \eta_l^\dag$ ($\hat \eta_l$) denoting the creation (annihilation) operator for a single-particle Anderson eigenstate $\phi_l$ with eigenvalue $\epsilon_l$.
As a second step, we express $\hat{\mathcal{V}}=\sum_{lmnk} \mathcal{B}_{lmnk} \hat \eta_l^\dag \hat \eta_m  \hat\eta_n^\dag \hat \eta_k $ in terms of the $\{\hat \eta_l\}$. 
Then we neglect all contributions that do not commute with the $\{\hat I_l\}$ so that we arrive at the following desired $l$-bit Hamiltonian:
\begin{equation} 
 \hat{H}^{\text{eff}} = \sum_l \epsilon_l \hat{\eta}^\dagger_l \hat{\eta}_l + \sum_{l,m} \mathcal{B}_{l,m} \hat{\eta}^\dagger_l \hat{\eta}_l \hat{\eta}^\dagger_m \hat{\eta}_m \, ,
 \label{eq:effective}
\end{equation}
with $\mathcal{B}_{l,m}=\mathcal{B}_{llmm}-\mathcal{B}_{lmlm}$.
% can be computed, as we show explicitly for one example below, so that we have all the necessary microscopic information for the targeted calculations.
%
This construction relies on the perturbative nature of an MBL phase, in which the integrals of motion of the system $\{\hat{I}_l\}$  can be obtained perturbatively from the non-interacting ones $\{\hat{\eta}_l^\dagger \hat{\eta}_l \}$~\cite{Basko06,doi:10.1002/andp.201600322,PhysRevB.97.060201, Ros15, Rade16}. Thus, as a first approximation in the limit of weak-interactions, the integrals of motion can be taken as the ones of the non-interacting case.   
In the concluding discussion we will outline how one can improve systematically the accuracy of the $l$-bits by accounting for higher orders $V$~\footnote{See Supplemental Material for a statistical analysis of the discarded elements $\mathcal{B}_{l,m,n,k}$}.
For the following, we will use the representation above and show that it is already sufficient to capture quantitatively the dynamics for small $V$.

\begin{figure}[t]
\includegraphics[width=1.\columnwidth]{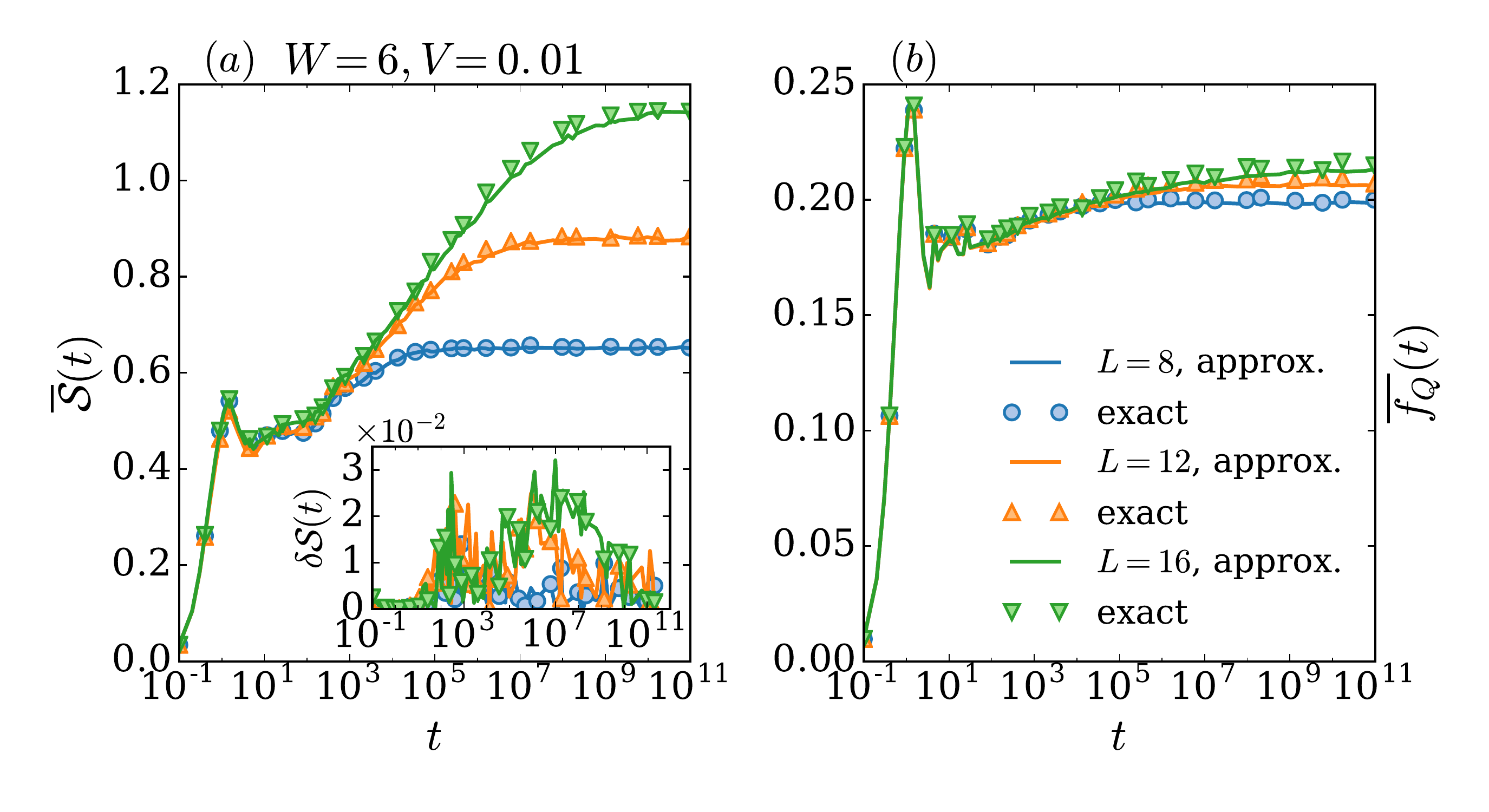}
\caption{(a): Bipartite half-chain entanglement entropy $\overline{S}(t)$ after a global
quantum quench for several systems sizes $(L)$ in $1D$. $\overline{S}(t)$ has been calculated using the exact
Hamiltonian $\hat{H}$ (exact) and the effective model $\hat{H}^{\text{eff}}$ (approx.).
The insect shows the relative error $\delta \mathcal{S}(t) = \overline{|\mathcal{S}(t)-\mathcal{S}^{\text{approx}}(t)|}/\overline{\mathcal{S}}(t)$, between the entanglement entropy calculated with  $\hat{H}$ the one calculated with $\hat{H}^{\text{eff}}$.
(b): QFI for the $1D$ MBL system for several system sizes compared with exact results.}
\label{fig:Com}
\end{figure}

Having discussed the construction of the $l$-bit Hamiltonian, we now outline how this can be used to study dynamics, which is based on two main properties.
\begin{figure}[t]
\includegraphics[width=1.\columnwidth]{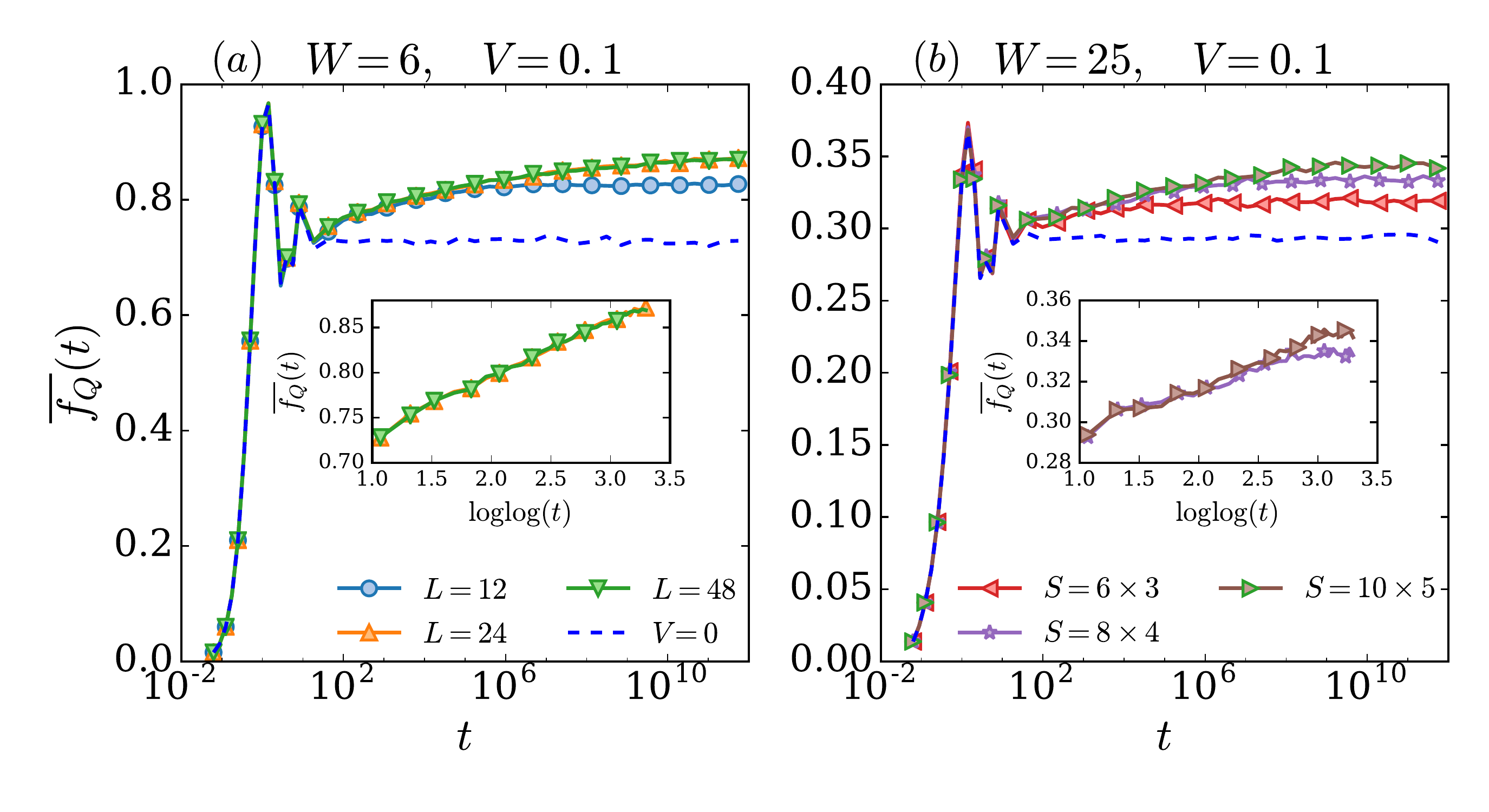}
\caption{(a): Disorder averaged QFI-density ($\overline{f_Q}(t) =\overline{\mathcal{F}_Q}(t)/2N$) 
for the $1D$ MBL model for several system sizes ($L$) and a fixed disorder and interaction strength. The inset shows $\overline{f_Q}(t)$ in a suitable scale 
to underline that $\overline{f_Q}(t)\sim \log{\log{t}}$. 
(b):  Disorder averaged QFI-density ($\overline{f_Q}(t)$) for the $2D$ model for several system sizes ($S$) and a fixed disorder and interaction strength. The inset shows that also in this case $\overline{f_Q}(t)\sim \log{\log{t}}$. 
For both panels the evolution has been obtained using $\hat{H}^{\text{eff}}$. Dashed-lines are for the non-interacting case ($V=0$) for the largest system size in each panels.}
\label{fig:Fig1}
\end{figure}
First, the time evolution of $\hat \eta_l$ and $\hat \eta_l^\dag$ can be determined analytically via $\hat{\eta}_l^\dagger(t) = \exp[i t\epsilon_l  +it\sum_m \tilde{\mathcal{B}}_{l,m} \hat \eta^\dagger_m \hat \eta_m] \hat \eta^\dagger_l$
where $\tilde{\mathcal{B}}_{l,m} = \mathcal{B}_{m,l} +  \mathcal{B}_{l,m}$.
\begin{figure*}[t]
\includegraphics[width=1.\textwidth]{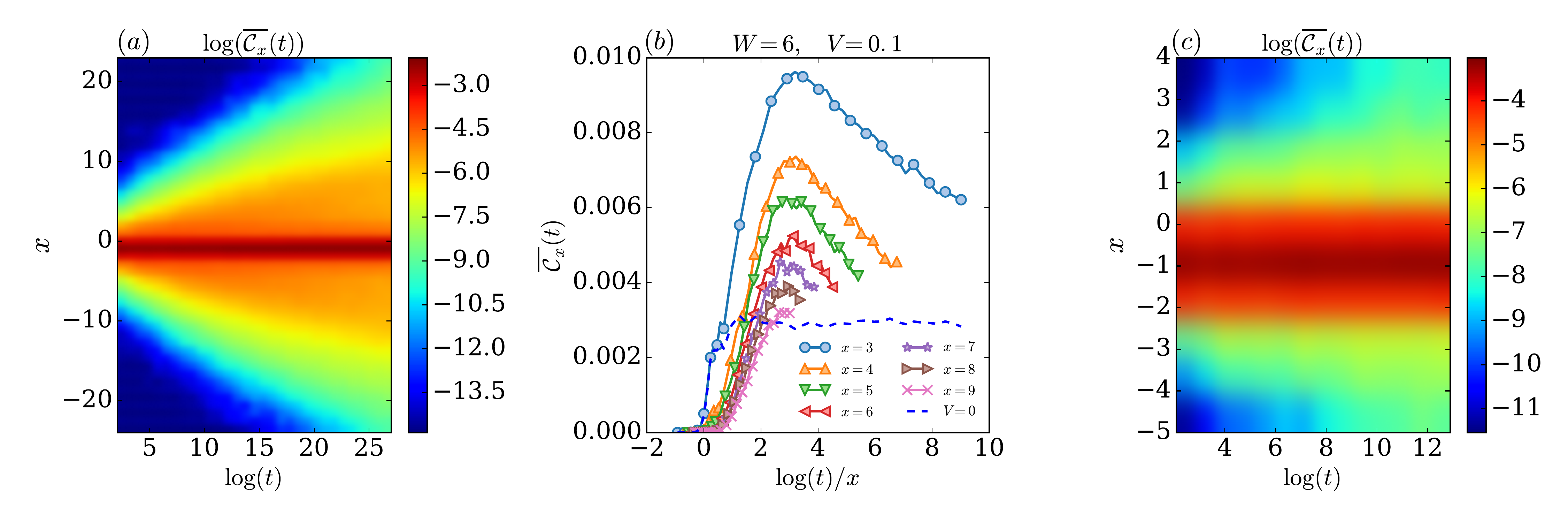}
\caption{(a): Logarithmic light-cone 
 calculated using the $\mathcal{C}_x(t)$ for the effective model in $1D$. 
 (b): $\mathcal{C}_x(t)$ as a function of time, the time has been 
 properly rescaled to get the collapse of the curves. It also shows the non-interacting case ($V=0$) (dashed-line). For both panels (a) and (b) $L=48$. 
 (c): Logarithmic light-cone  for the $2D$ case with $S=10\times 5$ in the $x$-direction calculated using $\mathcal{C}_{x,0}(t)$ evolved with the effective model with $W=25$ and $V=0.1$.}
\label{fig:Fig2}
\end{figure*}
Second, for an initial state $|\psi \rangle$, which is a product states in terms of the bare fermions, i.e. Gaussian, the expectation values of time-evolved local 
observables and correlation functions can be reduced to the evaluation of Slater determinants, which can be done very efficiently. For example, for a generic local observable 
$\hat{A} = \sum_{l,m} a_{l,m} \hat \eta_l^\dag \hat \eta_m$, we need only to calculate $\langle \hat{\eta}^\dagger_l \hat{\eta}_m (t)\rangle =  e^{it(\epsilon_l - \epsilon_m)} \langle \hat{\eta}_l^\dagger e^{it \sum_p(\tilde{\mathcal{B}}_{l,p} -\tilde{\mathcal{B}}_{p,m}) \hat{\eta}^\dagger_p\hat{\eta}_p} \hat{\eta}_m \rangle$,
where $\langle \dots \rangle = \langle \psi | \dots | \psi \rangle$. The term $\langle \hat{\eta}_l^\dagger e^{it \sum_p(\tilde{\mathcal{B}}_{l,p} -\tilde{\mathcal{B}}_{p,m}) \hat{\eta}^\dagger_p\hat{\eta}_p} \hat{\eta}_m \rangle$ can be 
efficiently computed using Wick's theorem~\cite{Peskin:1995ev}, interpreting $e^{it \sum_p(\tilde{\mathcal{B}}_{l,p} -\tilde{\mathcal{B}}_{m,p}) \hat{\eta}^\dagger_p\hat{\eta}_p} $ as an effective time-evolution operator of the quadratic Hamiltonian $\hat{H}^{(l,m)} = \sum_p(\tilde{\mathcal{B}}_{l,p} -\tilde{\mathcal{B}}_{m,p}) \hat{\eta}^\dagger_p\hat{\eta}_p$.
Importantly, such initial conditions are typical choices in theory~\cite{Bar12, Ser14, Russo18, Vass15, She17,Dogg18,Pro08,Raj16} and have been realized in the MBL context experimentally~\cite{Lus17,schreiber15,Choi16,Bordia16}.
%
%The knowledge of the time dependence  of the operators $\{\eta_l^{\dagger}\}$  allow us to calculate efficiently (polynomially in system size) the time expectation value of local observables for large times, making use of free-fermions techniques~\cite{1751-8121-42-50-504003}. 

For concreteness, we demonstrate our method for the Hamiltonian~\cite{Bera17, Luitz15, Pal10, Proz08, Bera15} 
\begin{equation} 
 \hat{H} :=-\frac{1}{2}\sum_{\langle\textbf{i},\textbf{j} \rangle} (\hat{c}^\dagger_{\textbf{i}} \hat{c}_{\textbf{j}} + h.c.)  + \sum_{\textbf{j}}  h_\textbf{j} \hat{n}_\textbf{j} + V\sum_{\langle \textbf{i},\textbf{j}\rangle }  \hat{n}_\textbf{i} \hat{n}_{\textbf{j}} 
\label{eq:Ham}
\end{equation}
where $\hat{c}_\textbf{j}^\dagger~(\hat{c}_\textbf{j})$ is the fermionic creation (annihilation) operator at
site $\textbf{j}$ and $\hat{n}_\textbf{j} = \hat{c}_\textbf{j}^\dagger \hat{c}_\textbf{j} $.
%$\langle \textbf{i}, \textbf{j}\rangle$ indicate summing over the first-neighbors.  
$\{h_\textbf{j} \}$  are random fields uniformly distributed between $[-W,W]$,  and $V$ is the interaction strength. 
We study the system both in a $1D$ lattice of size $L$ with periodic boundary conditions and defined in a rectangular lattice ($2D$) of size $S = L \times \frac{L}{2}$  
with periodic and open boundary conditions respectively in the $x$ and in the $y$ direction. We focus on half-filling $N/L=1/2$ ($N/|S|=1/2$) with $N$ the number of fermions. 
The $1D$ system is believed to have an MBL-phase at strong-disorder~\cite{Giu17,Bera15, Bera16,Pal10,Serb15, Luitz15}. 
The $2D$  case on the other hand has largely remained elusive due to the lack of efficient methods to simulate sufficiently large system sizes.
A recent experiment has given evidence of an MBL phase in a bosonic $2D$ system~\cite{Choi16}. Nevertheless, it is currently under debate whether MBL can be stable at all in $2D$~\cite{Choi16,Bordia16,Wahl17,Deroeck17,PhysRevB.99.205149,2018arXiv181104126K, 2019arXiv190204091T}.
The proposed mechanism for the breakdown of MBL relies on rare resonances which, however, only manifest on very long time scales, below which our $l$-bit description Eq.~\ref{eq:effective} could still be accurate at intermediate time scales.
%
%In our concluding discussion 
%While the existence of MBL in $2D$ is under debate on general grounds~\cite{Wahl17,Choi16, Bordia16,Deroeck17, Poti18}, a recent experiment with gives indication that an MBL phase might exist~\cite{Choi16}.
%Nevertheless, it has also been argued that rare regions with weak disorder could spoil localization, but only on extremely large time-scales~\cite{Deroeck17}, 
%making although also intermediate time scales as well as relevant. 

Following our prescription outlined before, we first diagonalize the noninteracting model by introducing $\hat \eta_l^\dagger = \sum_{\textbf{i}} \phi_l(\textbf{i}) \, \hat c^\dagger_\textbf{i}$.
This then leads to $\mathcal{B}_{l,m} = V \sum_{\langle \textbf{i}, \textbf{j}\rangle } [ |\phi_l(\textbf{i})|^2  |\phi_m(\textbf{j})|^2  - \phi_l(\textbf{i}) \phi_m (\textbf{i}) \phi_l (\textbf{j}) \phi_m (\textbf{j})  ]$.
In the remainder, we choose staggered initial states of charge-density type both for $1D$ $|\psi \rangle = \prod_{s=-L/4}^{L/4-2} c^\dagger_{2s} |0\rangle $ and 
for $2D$ $|\psi \rangle = \prod_{y=-L/2}^{L/2-1} \prod_{x=-L/4}^{L/4-2} c^\dagger_{(2x,y)} |0\rangle$, motivated by recent experiments~\cite{Choi16}.
Disorder averaged quantities will be indicated with an overline, e.g. $\overline{\langle \hat{n}_{\textbf{i}} \rangle}$.
%where $\{\phi_l\}$ are the exponentially localized single-particle eigenstates.
%After a suitable relabeling of the indexes $(l, m)$, $\mathcal{B}_{l,m}\sim V e^{-|l-m|/\xi_{\text{loc}}}$, where $\xi_{\text{loc}}$ is the single particle localization length.

{\it Benchmark for quantum-information propagation}---
We now compare the exact dynamics of $\hat{H}$ with the one generated by $\hat{H}^{\text{eff}}$.
For the benchmark we choose to study quantum information (entanglement) propagation which inherits one of the central and nontrivial features of MBL phases.
In Fig.~\ref{fig:Com} we show data for two measures both obtained using exact diagonalization and via our effective Hamiltonian~\footnote{For additional comparing data see the Supplemental material.}.
%Due to the limitation of exact diagonalization, we focus our comparison only on the 1D system in the limit of strong disorder ($W=6$) and weak interaction ($V = 0.01$)
First, this includes the half-chain entanglement entropy 
\begin{equation}
\mathcal{S}(t)=-\text{Tr} \hat{\rho}_{L/2} (t) \log{\hat{\rho}_{L/2}(t)}, 
\end{equation}
where $\hat{\rho}_{L/2}(t)$ denotes the  reduced  density  matrix  of half of the system. Second, we study the quantum Fisher information (QFI) related to the initial charge-density pattern defined by
\begin{equation}
\label{eq:QFI}
 \mathcal{F}_Q(t) = 4 \left[ \langle \hat{\mathcal{O}}(t)^2 \rangle - \langle \hat{\mathcal{O}}(t) \rangle^2 \right], \quad \hat{\mathcal{O}} = \sum_x (-1)^x \hat n_x .%=  \frac{1}{2L} \sum_{x,x1} (-1)^{x+x1} \langle \hat{n}_x \hat{n}_{x1} \rangle -  \frac{1}{2L}\left ( \sum_x (-1)^x \langle  \hat{n}_x \rangle \right )^2.
\end{equation}
The QFI probes the propagation of quantum correlations and is an entanglement witness~\cite{Hauke2016,Smith2016, Petz,PhysRevLett.72.3439, 2014Strobel, 2012Toth, 2012Hyllus}, that has been also measured experimentally in the MBL context~\cite{Smith2016}.
%Moreover, QFI is  proportional to the variance of the Hamming distance ($\propto \sum_i \langle   \hat{n}_i(t) \hat{n}_i(0) \rangle$), 
%which gives information on the spread of the evolved state in the Fock space~\cite{Smith2016}.
%Recently, $\mathcal{F}_Q(t)$ has been experimentally measured for a spins MBL-model with long-range hopping, probing a difference in its time dependence between an interacting and a non-interacting localized phase.
%From now on, the initial state $|\psi \rangle$ for the $1D$ model is taken to be $\prod_{s=-L/4}^{L/4-2} c^\dagger_{2s} |0\rangle $ (charge-density state) and
%the average over disorder will be indicated with an overline, e.g. $\overline{\langle \hat{n}_x \rangle}$.

As we can see from Fig.~\ref{fig:Com}(a) the effective model reproduces not only qualitatively the unbounded logarithmic growth of the entanglement entropy~\cite{Bar12,Aba13}, but even more importantly also 
quantitatively correctly in the long-time limit. In particular, the inset in Fig.~\ref{fig:Com}(a) shows that the relative error $\delta \mathcal{S}(t) = \overline{|\mathcal{S}(t)-\mathcal{S}^{\text{approx}} (t)|}/\overline{\mathcal{S}}(t)$ 
is a bounded function of time and remains smaller than 3$\%$ for all times.
Let us note that the results for $\delta \mathcal{S}(t)$ gives evidence that our method not only reproduces the logarithmic growth after disorder averaging but even for individual random configurations.
%
%We computed $\mathcal{S}(t)$ using our method by reconstructing the full reduced density matrix $\hat{\rho}_{L/2}$, which is not scalable for large systems and is used here only for the benchmark against exact diagonalization.
%
Similarly, also for the QFI the dynamics generated by the effective Hamiltonian follows closely the exact one, see Fig.~\ref{fig:Com}(b) where we define the QFI-density $f_Q = \mathcal{F}_Q/2N$~\cite{Smith2016}.
While the entanglement entropy serves as a prime example for MBL properties, its computation within our method is not scalable to large system sizes.
This, however, is different for the QFI which can still be computed efficiently even for large systems, which allows us to also access it in $2D$, see below.
It is important to note, that our method reproduces the exact dynamics also for times longer than the naively expected range of validity of perturbation theory $(\sim 1/V)$, what can be understood from a statistical analysis of the discarded elements $\mathcal{B}_{l,m,n,k}$~\footnote{See Supplemental Material for a statistical analysis of the discarded off-diagonal elements $\mathcal{B}_{l,m,n,k}$}.  
%and
%\begin{equation}
%\langle \hat{n}_x \rangle_{\text{time ave.}} = \lim_{T\rightarrow \infty} \frac{1}{T} \int_0^T ds \langle \hat{n}_x \rangle (s),
%\end{equation}
%The average over disorder is indicated with an overline, e.g. $\overline{\Delta n^2_x(t)}$.
%First, we compare the exact dynamics performed using $\hat{H}$ with the one using $\hat{H}^{\text{eff}}$. We show that in the limit of strong localization and weak interaction our method reproduce the exact dynamics for longer times much beyond  
%than what perturbation theory crudely predict ($t\sim 1/V$). 
%Second, we second simulate the listed quantities using the effective Hamiltonian Eq.~\eqref{eq:effective}, reaching time scale for larger system size that so far have not been observed, we confirm many of the dynamical properties in an MBL phase 
%(e.g. logistically growth of the entanglement entropy). Furthermore, it allows to underline new properties (e.g. QFI has a $\log \log (t)$ propagation) and to tackle the 2D case.  

{\it Results}---
Having shown that our method reproduces quantitatively the exact dynamics at a controlled error, we now aim to further demonstrate the capabilities of our method. 
We target this goal by addressing several aspects of MBL systems which up to now have not been accessible or could not be settled due to system size limitations.
This includes aspects of quantum information propagation, logarithmic spread of correlations, and temporal fluctuations of local observables both in $1D$ and $2D$.
In the following, we choose a larger interaction strength $V=0.1$ instead of $V=0.01$ as used for Fig.~\ref{fig:Com}, which increases slightly the relative error in the computed quantities, but on the same time allows us to amplify the influence of interaction effects.
%All the considered quantities can be used as dynamical indicators to distinguish an interacting from a non-interacting localized phase.

%Moreover, the initial state $| \psi \rangle$ for the $2D$ model has been taken equal to $|\psi \rangle = \prod_{y=-L/2}^{L/2-1} \prod_{x=-L/4}^{L/4-2} c^\dagger_{(2x,y)} |0\rangle$. 

Figure \ref{fig:Fig1} shows $f_Q(t)$ for the $1D$ (a) and the $2D$ case (b), respectively, now computed for much larger systems than done for the benchmark in Fig.~\ref{fig:Com}.
For the $2D$ model we choose the QFI along the x-direction, i.e., $\hat{\mathcal{O}} = \sum_{x} (-1)^{x} \hat n_x$ with $\hat{n}_x := \hat{n}_{(x,0)}$.
For comparison we also include the results for the noninteracting models, which show quick saturation to a system-size independent value.
For nonvanishing interactions, the behavior of $\overline{\mathcal{F}_Q}(t)$ changes completely and we observe a slow growth, which is consistent with $\overline{\mathcal{F}_Q}(t)\sim \log{\log{t}}$ (insets) 
over many decades in time and almost independent of system size.
As a consequence, we are capable to demonstrate slow quantum information propagation in $2D$ MBL systems, which up to now has not been possible by other methods~\cite{2019arXiv190204091T,PhysRevB.97.060201, 2018arXiv181104126K, Wahl17}.
%
%giving us the possibility to extract the extremely slow propagation $\log{\log{t}}$. 
%In a recent experiment in trapped ions using a long-range disordered Ising model is 
%shown that $\overline{\mathcal{F}_Q_Q}(t)\sim \log{t}$~\cite{Smith2016}. This discrepancy could be due to the presence of long-range hopping,
%
%\comment{Maybe we can put the following somewhere else? which could modify the type of dephesing mechanisms leading to $\mathcal{B}_{0,l} \sim 1/l^{\beta}$\footnote{G. De Tomasi in preparation}}.
In a recent experiment in trapped ions implementing a long-range disordered Ising model evidence for an intermediate $\overline{\mathcal{F}_Q}(t)\sim \log{t}$ growth has been found~\cite{Smith2016}, which, however, might be due to the fact that the system could be in an algebraic MBL phase~\cite{PhysRevB.99.054204, 2018arXiv181009779B}, leading to $\mathcal{B}_{0,l} \sim 1/l^{\beta}$ with power-law instead of exponential dependence~\cite{PhysRevB.99.054204, 2018arXiv181009779B,PhysRevA.99.033610, 2019arXiv190502286M}.

As a next step we aim at studying quantum correlation spreading via the two-point connected correlation function, defined by
%%%%%%%%%%%%%%%%%%
\begin{equation}
 \mathcal{C}_x(t) = | \langle \hat{n}_x(t) \hat{n}_0 (t)\rangle -   \langle \hat{n}_x (t)\rangle \langle \hat{n}_0 (t)\rangle | \, .
\label{eq:C}
 \end{equation}
 $\mathcal{C}_x(t)$ has been used in several quantum systems~\cite{Cheneau2012, PhysRevA.89.031602, L_uchli_2008, PhysRevB.97.205103, PhysRevLett.113.187203, PhysRevB.96.064303} to quantify the time $t$ required to correlate two sites at some distance $x$, giving rise to the so called light-cone of propagation of correlations.
Moreover, $\mathcal{C}_x(t)$ has been measured in a recent experiment in a disordered Bose-Hubbard chain to probe the existence of an MBL-phase~\cite{2018arXiv180509819L}.
The $1D$ case we address in Fig.~\ref{fig:Fig2}(a), where we show a color plot of $\mathcal{C}_x(t)$ displaying the logarithmic light-cone~\cite{2018arXiv180509819L, Cheneau2012, PhysRevA.89.031602, L_uchli_2008, PhysRevB.97.205103, PhysRevLett.113.187203, PhysRevB.96.064303, Deng17, 2018arXiv180706086S} over many decades with quantum correlations  spreading  in  space  only  logarithmically  slowly in time. 
\begin{figure}
	\includegraphics[width=1.\columnwidth]{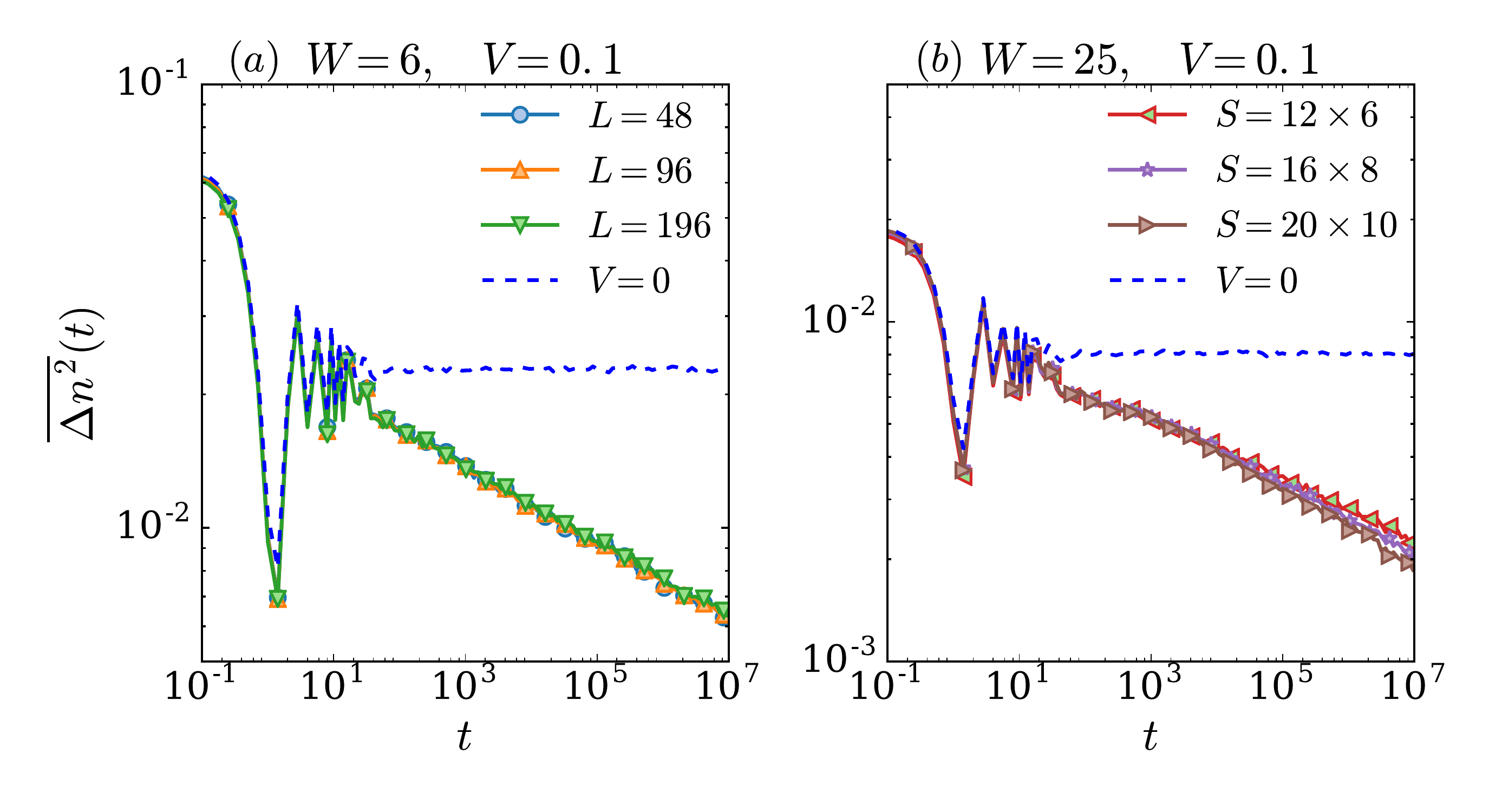}
	\caption{(a): Disorder average time fluctuations ($\overline{\Delta n^2}(t)$) for several system sizes $L$ for the $1D$ case, $\overline{\Delta n^2}(t)\sim t^{-\alpha}$. 
		(b): $\overline{\Delta n^2}(t)$ for several system sizes $S$ for the $2D$ case, also in this case its decay is consistent with an algebraic decay in time. 
		The non-interacting case ($V=0$) is also shown (dashed-lines) for the largest system size.
		For both cases the evolution has been performed using $\hat{H}^{\text{eff}}$.} 
	\label{fig:Fig3}
\end{figure}
Interestingly, however, we find that there exists a time scale $t_x^\star$ beyond which $\mathcal{C}_x(t)$ starts to decrease again, see Fig.~\ref{fig:Fig2} (b), an effect which has not yet been recognized before.
Remarkably, this indicates that quantum correlations are eventually scrambled in the long-time limit also in an MBL system, 
which might be consistent and even necessary with the expectation to reach in the long-time limit a state with volume-law entanglement entropy~\footnote{See Supplemental Material for a further analysis of the two-point correlation function $\mathcal{C}_x(t)$.}.
From the rescaling of the time axis used in Fig.~\ref{fig:Fig2}(b) we find evidence that this correlation time $t_x^\star$ scales exponentially with the distance $x$ ($\log{t_x^\star}\sim x$).
%Indeed, $t_x^\star$ is proportional to the time in which the site $x$ starts to be entangled with the site $x=0$, $t_x^\star \sim \tilde{\mathcal{B}}^{-1}_{0,x}\sim e^{x/\xi_\text{loc}}$. 
In the case of an Anderson insulator ($V=0$) quantum correlations are frozen in the long-time limit~\cite{Bar12, Aba13}, implying the saturation to non-zero value of $\mathcal{C}_x(t)$ (Fig. (b)~\ref{fig:Fig2} dashed-line). 
Finally, in Fig. (c)~\ref{fig:Fig2} we study correlation spreading in $2D$, where we found again, like in $1D$, the same logarithmically slow propagation.

As opposed to an Anderson insulator  it has been argued that an MBL system can show relaxation~\cite{Serb15, Poll16}, meaning that expectation values of local observables reach at long time a stationary value in the thermodynamic limit with decaying temporal fluctuations. 
Here, we use our method to reexamine the temporal fluctuations in $1D$ and to study them also for $2D$ systems.
These are defined for $\hat{n}_x$ via
\begin{equation}
 \Delta n^2(t) = \frac{1}{L} \sum_x \Delta n^2_x(t), \, \Delta n^2_x(t)  = \left ( \langle \hat{n}_x \rangle (t)  - \langle \hat{n}_x \rangle_{\text{tav} }   \right )^2 \,,
 \label{eq:time_flu}
\end{equation}
%where 
%\begin{equation}
%\Delta n^2_x(t)  = \left ( \langle \hat{n}_x \rangle (t)  - \langle \hat{n}_i \rangle_{\text{time ave.} }   \right )^2,
%\end{equation}
where $\langle \hat{n}_x \rangle_{\text{tav}}$ denotes the long-time average of $\langle \hat{n}_x \rangle (t)$.
As shown in Fig.~\ref{fig:Fig3}, both in $1D$ and $2D$ the temporal fluctuations exhibit an algebraic decay with time, $\overline{\Delta n^2}(t)\sim t^{-\alpha}$.
As a reference we have included also the data for the noninteracting cases ($V=0$, dashed-lines), where temporal fluctuations remain non vanishing for all times.
We find that the exponent $\alpha$ is proportional to the single-particle localization length $\xi_{\text{loc}}$~\footnote{For additional data see the Supplemental material.}, for which we now aim to give an analytical argument.
This shows that our method not only can be used for numerically computing quantities but also for analytical predictions.
%
%For $V=0$, in first approximation, $\Delta n^2_x(t)$ involves only single-particle eigenenergies with eigenfunctions such that their localization centers are around $x$, thus being a sum of finite number of oscillating terms and therefore does not decay. 
%For $V\ne 0$, $\Delta n^2_x(t)$ contains also the eigenenergies such that their localization centers are around $x$ but now these finite number of degrees of freedom  are exponentially weakly coupled also with all other eigenenergies due to the  dephasing terms $\{\mathcal{B}_{x,l}\}$. Thus, $\Delta n^2_x(t)$ contains an extensive number of oscillating terms which are not in phase to each other,
%thus producing its decay. 
For that purpose we consider a special initial state $|\psi \rangle = \prod_l^L \frac{\hat{\eta}_l + \hat{\eta}_l^\dagger}{\sqrt{2}} | 0 \rangle$ 
for which the calculations are simplified but which gives qualitatively the same decay of the temporal fluctuations~\footnote{For additional data see the Supplemental material.}.
For this state we find $\Delta n_x^2 (t) =  [ \sum_{l\ne m} \phi_l(x) \phi_m(x) e^{i(\epsilon_l-\epsilon_m)t} Q_{lm} ]^2 $ with $Q_{lm} = 2^{-2} \prod_{k=l+1}^m \sin( A_k^{l,m}t) \prod_{s\ne k}\cos(A_s^{l,m}t) $ and $A_k^{l,m} =  (V/2) (\tilde{\mathcal{B}}_{m,k} - \tilde{\mathcal{B}}_{l,k})\sim V e^{-\frac{\min(|m-k|, |l-k|)  }{\xi_\text{loc}}}$.
The sum over $(l,m)$ can be restricted only to eigenstates, whose centers are located within a distance $\xi_{\text{loc}}$ away from $x$. 
Each term of the $\cos$'s and $\sin$'s with argument $A_k^{l,m}$ decays exponentially in $k$, which leads to a power-law in time~\cite{She17} with an exponent proportional to $\xi_{\text{loc}}$, implying $\Delta n_x^2 (t) \sim t^{-c\xi_{\text{loc}}}$~\cite{Ser14}. 
%and perturbation 
%theory ensure that our approximation ($\hat{H}^{\text{eff}}$) is faithful for times order $\sim 1/V$.   

{\it Conclusions}---
In this work, we have formulated a method which allows to efficiently study the dynamics of weakly-interacting localized fermions.
The accuracy of the approach can be further increased systematically by taking into account those contributions to the interaction term, which are not commuting with the bare integrals of motion $\hat I_l$ and which have been completely neglected in the present study.
For example, to lowest order they can be accounted for by a Schrieffer-Wolff transformation.
Our method can be applied to any weakly interacting MBL system, which exhibits an $l$-bit representation, not only limited to the quantum quench dynamics studied here.
Thus, it can be used also to study, for example, also driven Floquet MBL systems~\cite{2017Bordia} such as they appear in discrete time crystals~\cite{2017Choi,2017Zhang}, MBL bosonic systems, algebraic MBL~\cite{PhysRevB.99.054204, 2018arXiv181009779B,PhysRevA.99.033610} and MBL weakly coupled with thermal baths~\cite{Wu_2019,PhysRevB.90.064203, PhysRevB.95.035132, PhysRevLett.116.237203}.
However, let us note that even in cases where an MBL phase might not be stable asymptotically for infinite system sizes and infinite times, our method might still provide a description on intermediate time scales (e.g. MBL in $2D$~\cite{Choi16,Bordia16,Wahl17,Deroeck17,PhysRevB.99.205149, 2018arXiv181104126K, 2019arXiv190204091T}).
Overall, our method maps the dynamical quantum many-body problem onto a system of classical degrees of freedom of mutually commuting operators,
similar in spirit to recent works where dynamical problems have been solved using classical~\cite{SciPostPhys.4.2.013} or artificial neural networks~\cite{2017Carleo}.
Instead of solving the problem in the basis of the bare particles, 
our work shows that a simple basis transformation onto more convenient degrees of freedom can improve the accuracy and efficiency dramatically, which might also be of relevance for the aforementioned approaches.

{\it Note}---Very recently the dynamics of one-point functions has been computed using a self-consistent Hartree-Fock method, which scales polynomially in system-size and time~\cite{PhysRevB.98.224205}. 

\begin{acknowledgments}
{\it Acknowledgments}---
We thank J.H. Bardarson, S. Bera, A. Burin, A. Eckardt, I. M. Khaymovich, M. Knap and D. Trapin  for several illuminating discussions. 
FP acknowledges the support of the DFG Research Unit FOR 1807 through grants no. PO 1370/2- 1, TRR80, the Nanosystems Initiative Munich (NIM) 
by the German Excellence Initiative, and the European Research Council (ERC) under the European Union's Horizon 2020 research and innovation program (grant agreement no. 771537). 
This research was conducted in part at the KITP, which is supported by NSF Grant No. NSF PHY-1748958. MH acknowledges support from the Deutsche Forschungsgemeinschaft via the
Gottfried Wilhelm Leibniz Prize program.
\end{acknowledgments}
\bibliography{Mean_bib}

%\acknowledgment
%{\it Acknowledgments}---

%{\it Acknowledgments}---
%We thank J.H. Bardarson, S. Bera, A. Burin, A. Eckardt, I. M. Khaymovich,  D. Trapin  for several illuminating discussions. 
%FP acknowledges the support of the DFG Research Unit FOR 1807 through grants no. PO 1370/2- 1, TRR80, the Nanosystems Initiative Munich (NIM) 
%by the German Excellence Initiative, and the European Research Council (ERC) under the European Union's Horizon 2020 research and innovation program (grant agreement no. 771537). 
%This research was conducted in part at the KITP, which is supported by NSF Grant No. NSF PHY-1748958. MH acknowledges support from the Deutsche Forschungsgemeinschaft via the
%Gottfried Wilhelm Leibniz Prize program.
\clearpage

\section{Supplemental material to Solving efficiently the dynamics of many-body localized systems at strong disorder}

{\it Using Free-Fermion Techniques}---
The effective model to describe an MBL-phase in the weak-interactions regime reads
\begin{equation} 
 \hat{H}^{\text{eff}} = \sum_l \epsilon_l \hat{\eta}^\dagger_l \hat{\eta}_l + \sum_{l,m} \mathcal{B}_{l,m} \hat{\eta}^\dagger_l \hat{\eta}_l \hat{\eta}^\dagger_m \hat{\eta}_m,
 \label{eq:effective}
\end{equation}
where $\hat{\eta}^\dagger_l = \sum_x \phi_l(\textbf{i}) \hat{c}^\dagger_x$ with $\{\phi_l\}$ and $\{\epsilon_l\}$ respectively  the single-particle wavefunctions and eigenvalues. The 
coefficient $\{\mathcal{B}_{l,m}\}$ are given by
\begin{equation}
 \mathcal{B}_{l,m} = V \sum_{\langle \textbf{i}, \textbf{j}\rangle } [ \phi_l(\textbf{i}) \phi_m (\textbf{i}) \phi_l (\textbf{j}) \phi_m (\textbf{j}) - |\phi_l(\textbf{i})|^2  |\phi_m(\textbf{j})|^2  ].
\end{equation}
\begin{figure}
\includegraphics[width=1.\columnwidth]{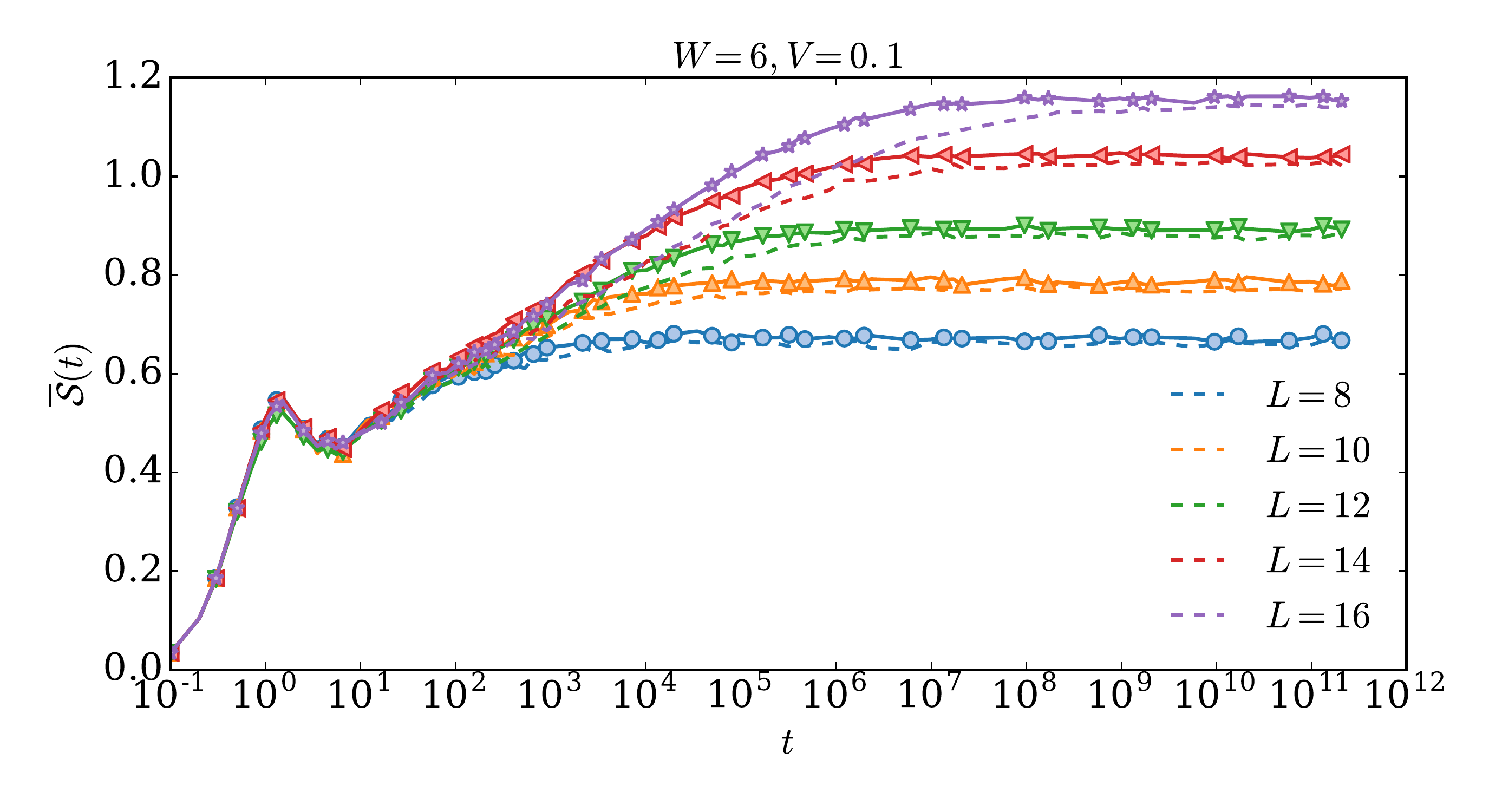}
\caption{$\overline{\mathcal{S}}(t)$ evolved with $\hat{H}$ and with $\hat{H}^{\text{eff}}$ (dashed-line) for the one-dimensional case for $W=6$, $V=0.1$ and several system sizes $L$.} 
\label{fig:S1}
\end{figure}
\begin{figure}
\includegraphics[width=1.\columnwidth]{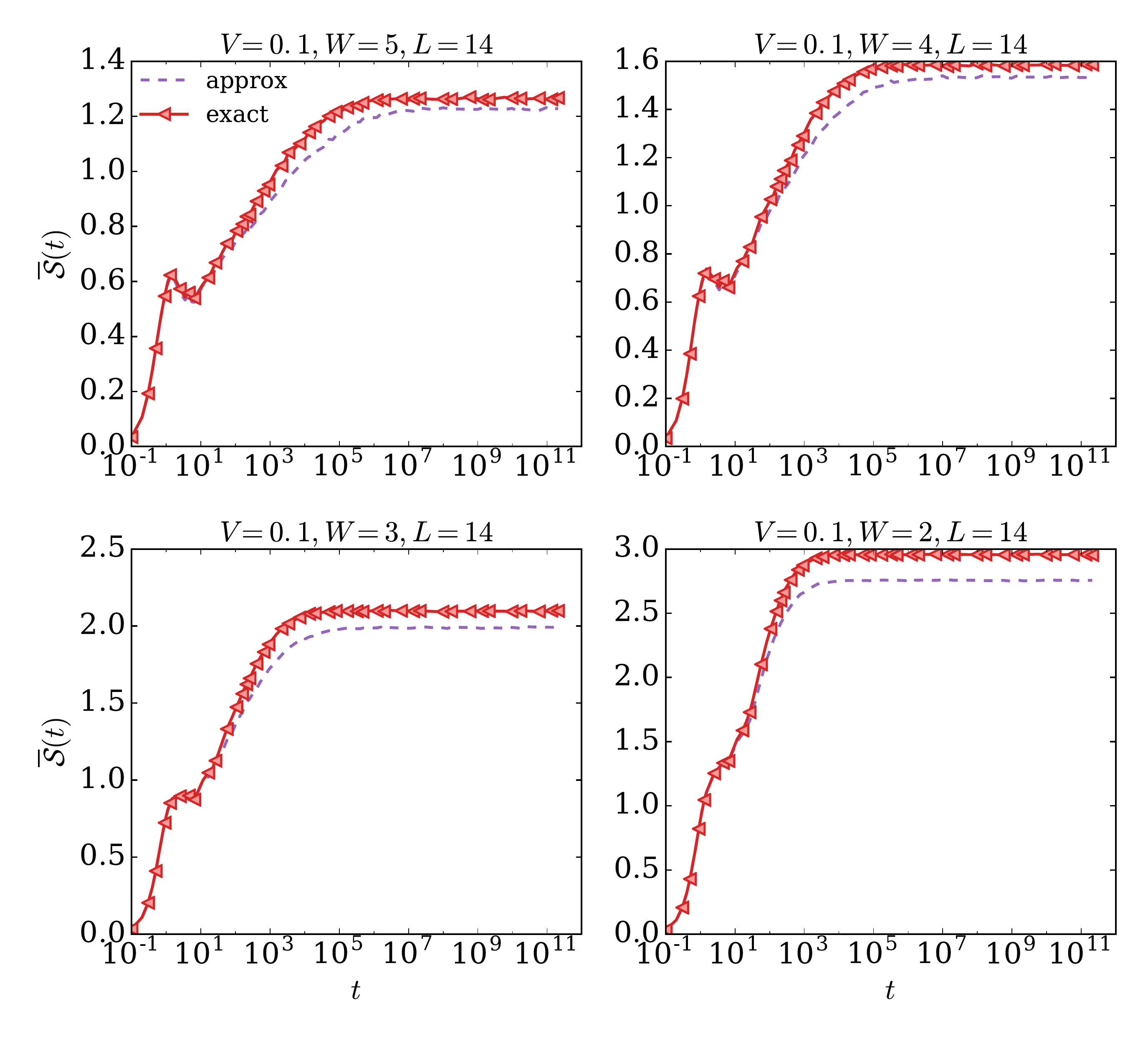}
\caption{$\overline{\mathcal{S}}(t)$ evolved with $\hat{H}$ and with $\hat{H}^{\text{eff}}$ (dashed-line) for fixed $L=14$ and $V=0.1$ and for several disorder strengths $W$.} 
\label{fig:S2}
\end{figure}
In this section, we give an example to show how to calculate efficiently the expectation values of local obeservables if the quantum evolution is performed using the Hamiltonian $\hat{H}^{\text{eff}}$. 
Let's consider the density-operator $\hat{n}_x = \hat{c}_x^\dagger \hat{c}_x$
\begin{equation}
\label{equation1}
 \langle \hat{n}_x \rangle = \sum_{l,m} \phi_l(x) \phi_m(x) e^{it(\epsilon_l - \epsilon_m)} \langle \hat{\eta}_l^\dagger e^{it \sum_p(\tilde{\mathcal{B}}_{l,p} -\tilde{\mathcal{B}}_{p,m}) \hat{\eta}^\dagger_p\hat{\eta}_p} \hat{\eta}_m \rangle,
\end{equation}
where $\tilde{\mathcal{B}}_{l,m} = \mathcal{B}_{l,m} + \mathcal{B}_{m,l}$. In Eq.~\ref{equation1} we have used the exact time-dependence of the operators $\{\hat{\eta}_l^\dagger\}$
\begin{equation}
  \hat{\eta}_l^\dagger(t) = e^{+i t\epsilon_l  +it\sum_m \tilde{\mathcal{B}}_{l,m} \hat{\eta}^\dagger_m \hat{\eta}_m} \hat{\eta}^\dagger_l.
\end{equation}
Now, the expectation value  $\langle \hat{\eta}_l^\dagger e^{it \sum_p(\tilde{\mathcal{B}}_{l,p} -\tilde{\mathcal{B}}_{m,p}) \hat{\eta}^\dagger_p\hat{\eta}_p} \hat{\eta}_m \rangle$ can be calculate 
with standard free-fermion technique~\cite{1751-8121-42-50-504003}, seeing the operator $e^{it \sum_p(\tilde{\mathcal{B}}_{l,p} -\tilde{\mathcal{B}}_{p,m}) \hat{\eta}^\dagger_p\hat{\eta}_p}$  a quantum-evolution operator for the quadratic Hamiltonian defined by
\begin{equation}
 \hat{H}^{(l,m)} = \sum_p(\tilde{\mathcal{B}}_{l,p} -\tilde{\mathcal{B}}_{m,p}) \hat{\eta}^\dagger_p\hat{\eta}_p.
\end{equation}
\begin{figure}
\includegraphics[width=1.\columnwidth]{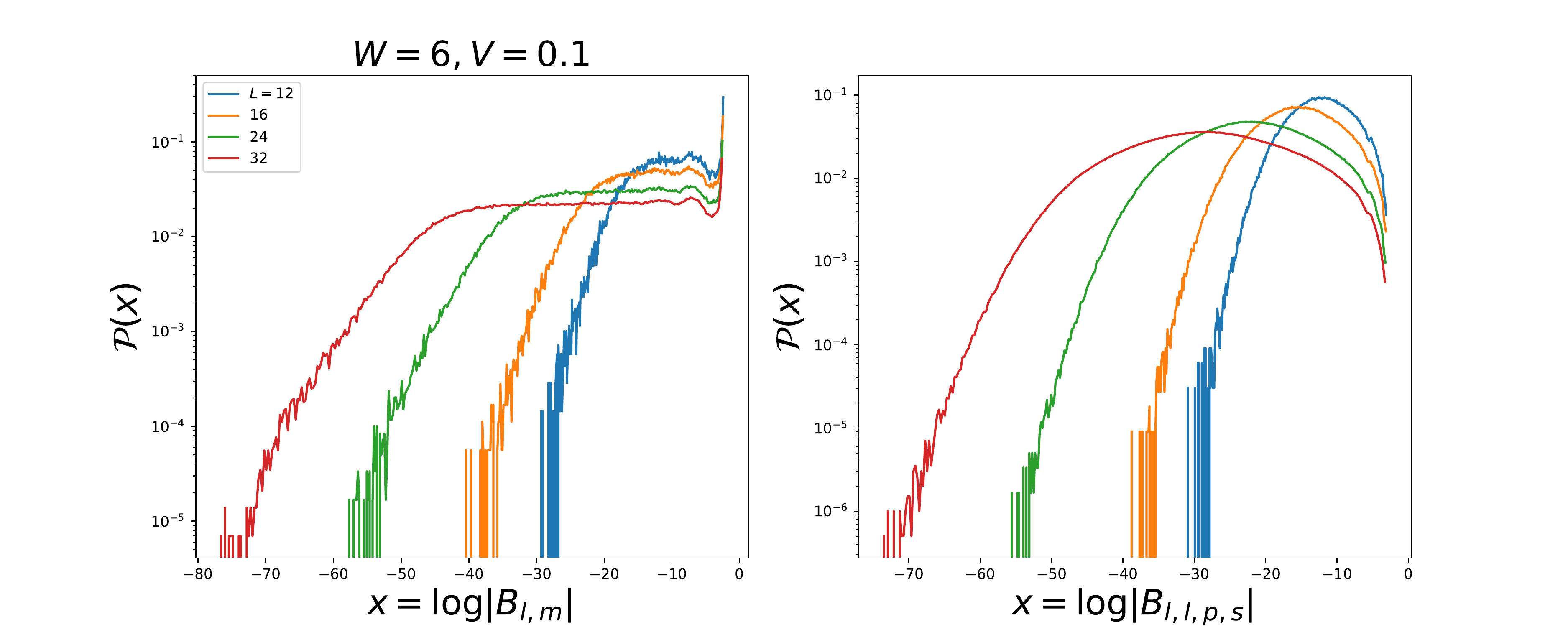}
\caption{The left-panel shows the probability distribution $\mathcal{P}(x)$ of $x=\log{|\mathcal{B}_{l,m}|}$ for several $L$ and $W=6$ and $V=0.1$. The right-panel shows $\mathcal{P}(x)$ for $x=\log{|\mathcal{B}_{l,l,p,s}|}$ for the same values of $L$, $W$ and $V$ as in the left-panel.}
\label{fig:S5}
\end{figure}
\begin{figure}
\includegraphics[width=1.\columnwidth]{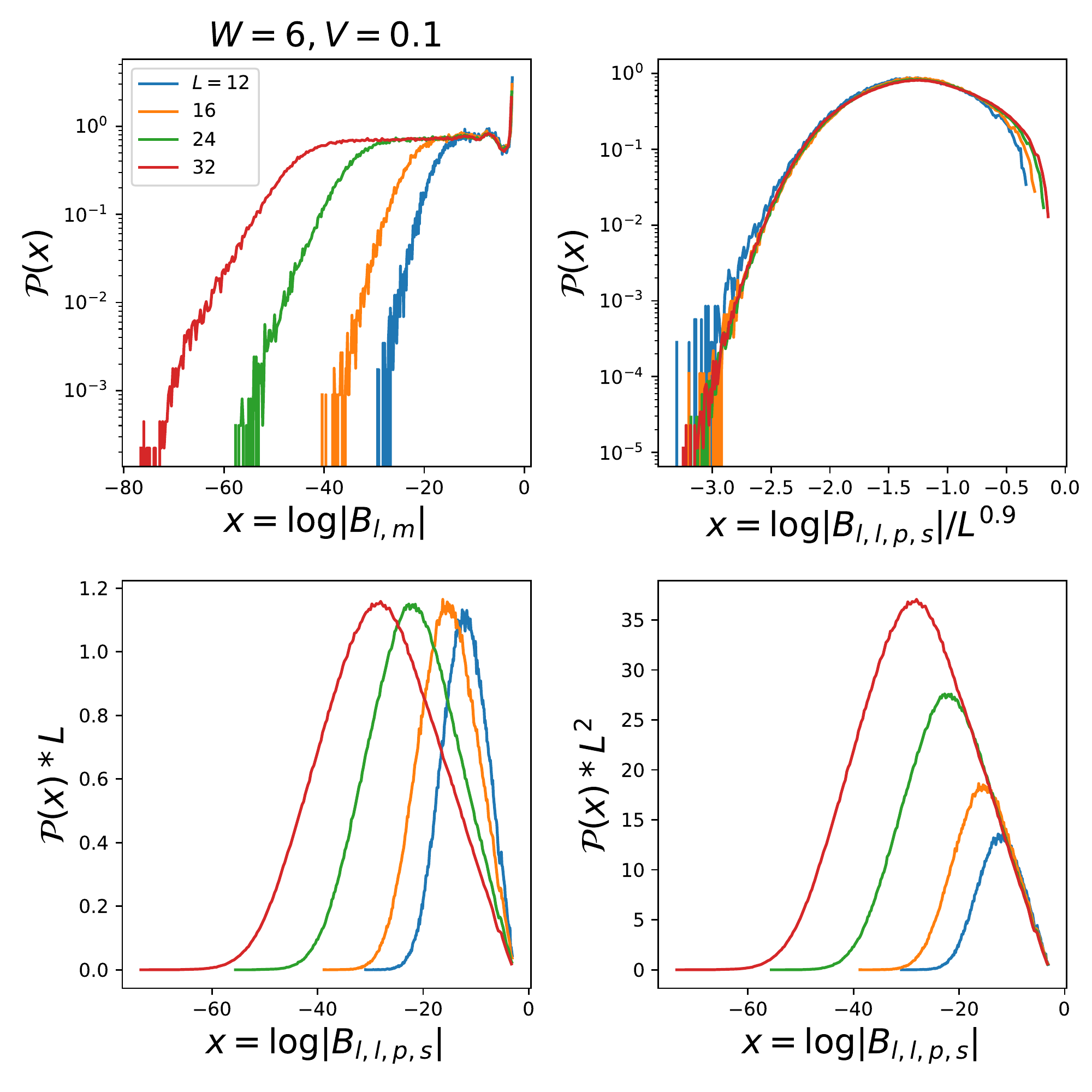}
\caption{The figure shows the rescaled probability distribution of $x=\log{|\mathcal{B}_{l,m}|}$ and $x=\log{|\mathcal{B}_{l,l,p,s}|}$ with $L$ to underline several characteristics. For $x=\log{|\mathcal{B}_{l,m}|}$ $\mathcal{P}(x\sim x_{\text{max}} )\le \mathcal{P}(x\sim 0)$ and in general  $\mathcal{P}(x\sim x_{\text{max}})\sim 1/L$. For $x=\log{|\mathcal{B}_{l,l,p,s}|}$ $\mathcal{P}(x\sim x_{\text{max}} ) \ge \mathcal{P}(x\sim 0)$, in particular  $\mathcal{P}(x\sim x_{\text{max}} )\sim 1/L$ and $\mathcal{P}(x\sim 0)\sim 1/L^2.$  } 
\label{fig:S6}
\end{figure}
\begin{figure}
\includegraphics[width=1.\columnwidth]{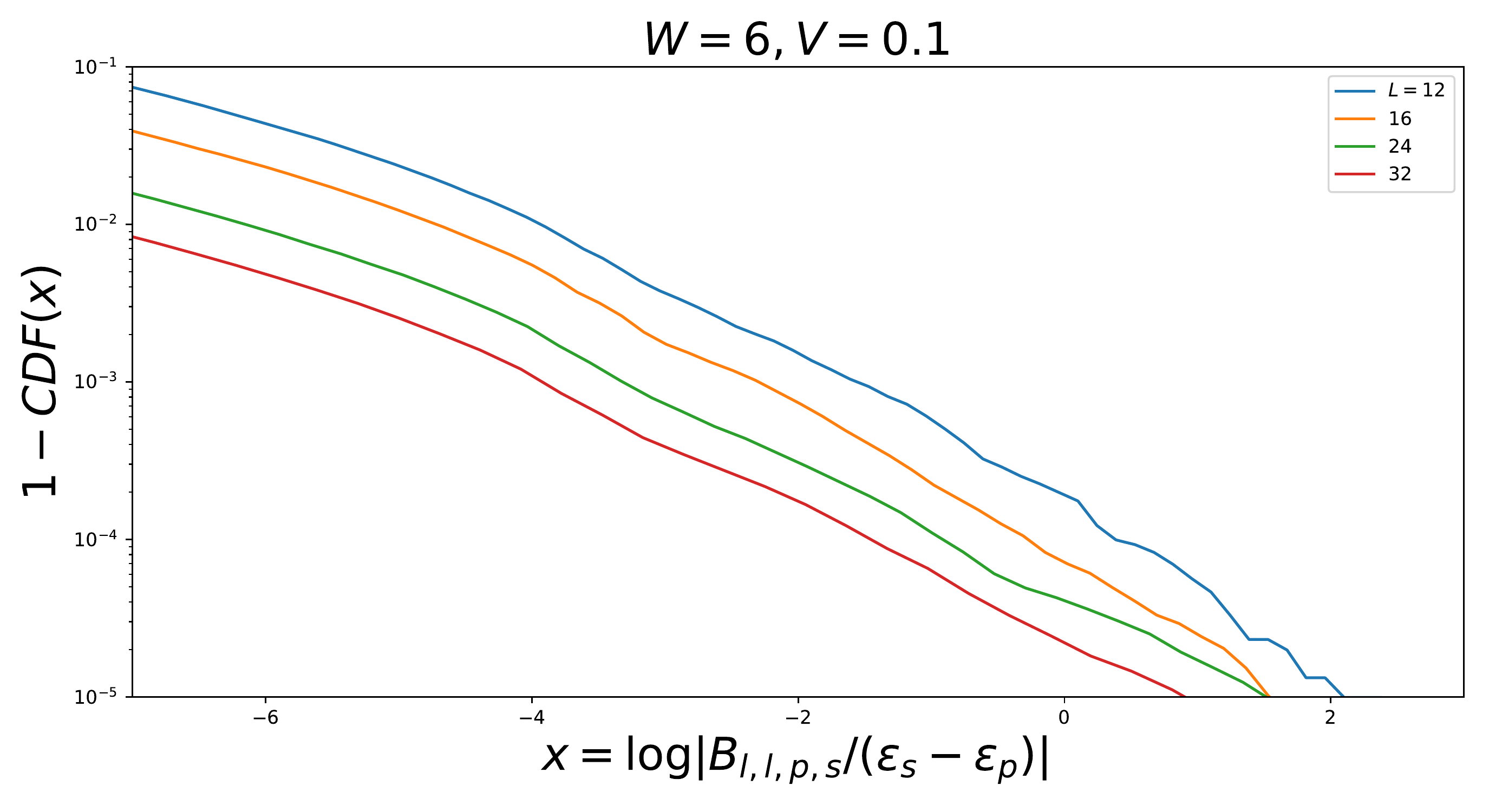}
\caption{The figure shows $1-CDF(x)$ for $x = \log{ \left |\frac{B_{l,l,p,s}}{(\epsilon_s - \epsilon_p)} \right |}$ for $W=6$, $V=0.1$ and several $L$. The probability of having resonances is $1-CDF(x\ge0)\sim10^{-4}$.} 
\label{fig:S7}
\end{figure}
\begin{figure}
\includegraphics[width=1.\columnwidth]{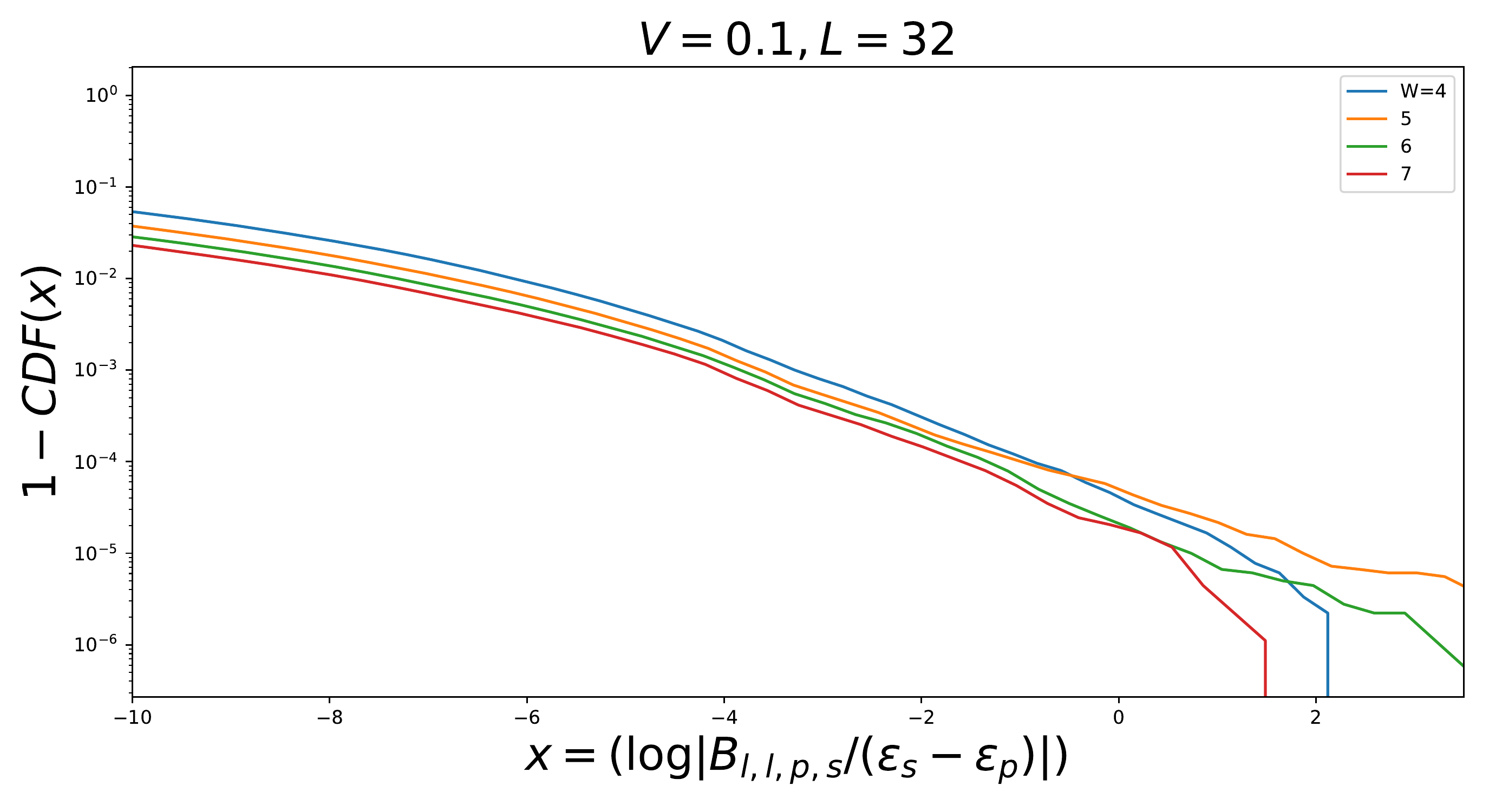}
\caption{The figure shows $1-CDF(x)$ for $x = \log{ \left |\frac{B_{l,l,p,s}}{(\epsilon_s - \epsilon_p)} \right |}$ for $L=32$, $V=0.1$ and several $W$.} 
\label{fig:S8}
\end{figure}
{\it Comparison with exact the exact results}---
Here, we show further data, comparing the quantum dynamics computed with the exact Hamiltonian $\hat{H}$ and the effective Hamiltonian $\hat{H}^{\text{eff}}$.
Figure~\ref{fig:S1} shows the entanglement entropy $\overline{\mathcal{S}}(t)$ for the one-dimensional system for $V=0.1$ and $W=6$, which are the values that have been used in the main text, for several system sizes $L$. 
$\mathcal{S}(t)$ independently if calculate with $\hat{H}$ or $\hat{H}^{\text{eff}}$ presents the typical log-growth propagation in an MBL-phase ($\sim \xi \log(t)$).
Although, the prefactor $\xi$ is different, since our approximation assumes that the localization length in 
the interacting case $(\xi)$ is the same as the non-interacting one $(\xi_{\text{loc}})$. In other words, for the approximated dynamics we have $\overline{\mathcal{S}}^{\text{approx}} (t) \sim \xi_{\text{loc}} \log{t}$, while for the exact one 
$\overline{\mathcal{S}}(t) \sim  \xi \log{t}$ with $\xi \sim \xi_{\text{loc}} + \mathcal{O}(V/W)$, making the relative error $\delta S(t)$ a bounded function of time. 
Figure~\ref{fig:S2} shows $\overline{\mathcal{S}}(t)$ for fixed interaction strength $V=0.1$ and system size $L=14$ for several disorder strengths $W$.
As expected our approximation works better for larger disorder strength, in any case independently of $W$ we have $\overline{S}^{\text{approx}}(t) \sim \log{t}$. 
A different approach to estimate the error done it in our approximation is to calculate the elements of matrix $\mathcal{B}_{l,m,n,p}$ that we have neglected. 
We will confine our discussion for the 1D case of the Hamiltonian studied in the main text. In this case the interactions elements in the Anderson basis are given by
\begin{equation}
 \mathcal{B}_{l,m,n,p} = V\sum_x \phi_l(x) \phi_m(x) \phi_n(x+1) \phi_p(x+1).
\end{equation}
There are two classes of elements that have not been considered
\begin{enumerate}
 \item \textit{Assistant hopping process} $\rightarrow$ $\mathcal{B}_{l,l,p,s}$ with $l\ne p\ne s$ (where two energy indexes are the same).
 \item $\mathcal{B}_{l,m,n,p}$ with $l\ne m, \ne n \ne p$ (where all indexes are different).   
\end{enumerate}
The element of matrix $\mathcal{B}_{l,m,n,p}$ involves the ``overlap'' between four localized wave functions, thus the second class of the neglected elements is significant smaller compared to the first one. We focus our attention on the assistant hopping terms ($\mathcal{B}_{l,l,p,s}$). 
Figure~\ref{fig:S5} shows the probability distribution $\mathcal{P}$ of $\mathcal{B}_{l,m}$ and of $\mathcal{B}_{l,l,p,s}$ over random configurations and energy indexes for $W=6$ and $V=0.1$ and several system sizes. It is possible to see that the probability distribution of $\mathcal{B}_{l,m}$ is picked for value of order one ($\mathcal{B}_{l,m}\sim \mathcal{O}(1)$), while most of them are equally distributed (plateau) to much smaller values,
\begin{equation}
 \mathcal{P}(x \sim x_{\textit{max}}) \le \mathcal{P}(x \sim 0) \quad x=\log{\mathcal{B}_{l,m}}, 
\end{equation}
where $x_{\textit{max}}$ is the value of $x$ for which $\mathcal{P}(x)$ has a maximum. 

Instead the probability distribution of $\mathcal{B}_{l,l,p,s}$ tend to zero for large values of $\mathcal{B}_{l,l,p,s}$, and most of its elements are distributed to much smaller values.
Thus contrarily to the previous case: 
\begin{equation}
 \mathcal{P}(x \sim x_{\textit{max}}) \ge \mathcal{P}(x \sim 0), \quad x=\log{\mathcal{B}_{l,l,p,s}}. 
\end{equation}
Moreover, in probability the elements $B_{l,m}$ dominate in magnitude the assistant hopping terms for these values of $W$ and $V$. 
\begin{figure}
\includegraphics[width=1.\columnwidth]{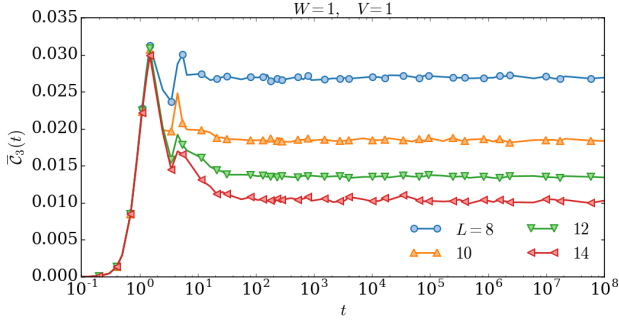}
\caption{$\mathcal{C}_x(t)$ with fix $x=3$ and several system sizes $L$ at weak disorder (ergodic phase). After a transient short time dynamic, $\mathcal{C}_x(t)$ saturates with time to a $L$-dependent value which scale exponentially fast to zero with $L$, $\lim_{t\rightarrow \infty} \mathcal{C}_x(t)\sim e^{-\alpha L}$.  } 
\label{fig:Sup_C}
\end{figure}
\begin{figure}
\includegraphics[width=1.\columnwidth]{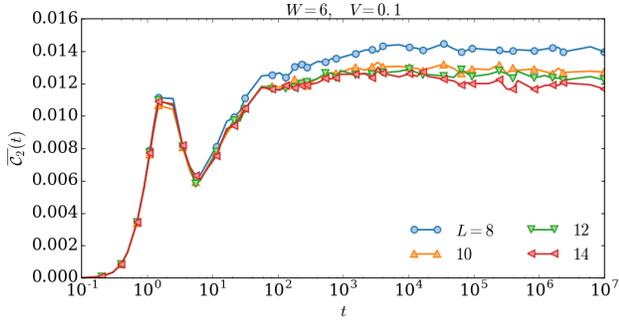}
\caption{$\mathcal{C}_x(t)$ with fix $x=2$ and several system sizes $L$ deep in the MBL phase ($W=6, V=0.1$). This results are obtained using exact diagonalization. Due to the limitation of system size the expected bending at large time for $\mathcal{C}_x(t)$ is only slightly visible.}
\label{fig:Sup_C1}
\end{figure}
\begin{figure}
\includegraphics[width=1.\columnwidth]{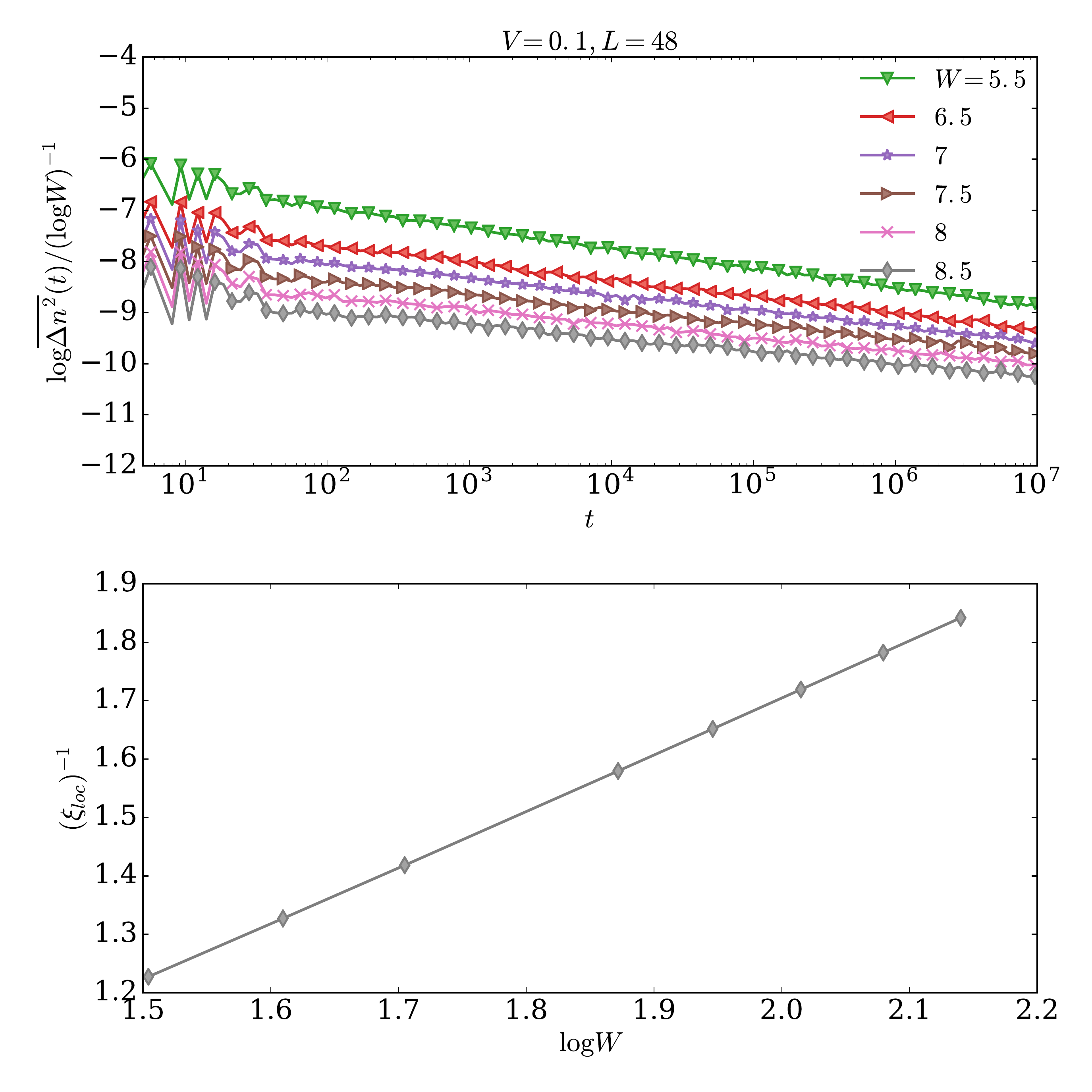}
\caption{The top-panel shows the time fluctuation $\overline{\Delta n^2}(t)$ calculated using $\hat{H}^{\text{eff}}$ for several $W$'s in the strong disorder limit and fixed $V$ and $L$. The initial state is given by 
$|\psi \rangle = \prod_{s=-L/4}^{L/4-2} c^\dagger_{2s} |0\rangle $ (charge-density state).
The bottom-panel shows $\xi_{\text{loc}}$ calculated with transfer-matrix technique at the band-center of the single-particle problem, giving evidence that $\xi_{\text{loc}} \sim \log{W}^{-1}$.} 
\label{fig:S3}
\end{figure}
\begin{figure}
\includegraphics[width=1.\columnwidth]{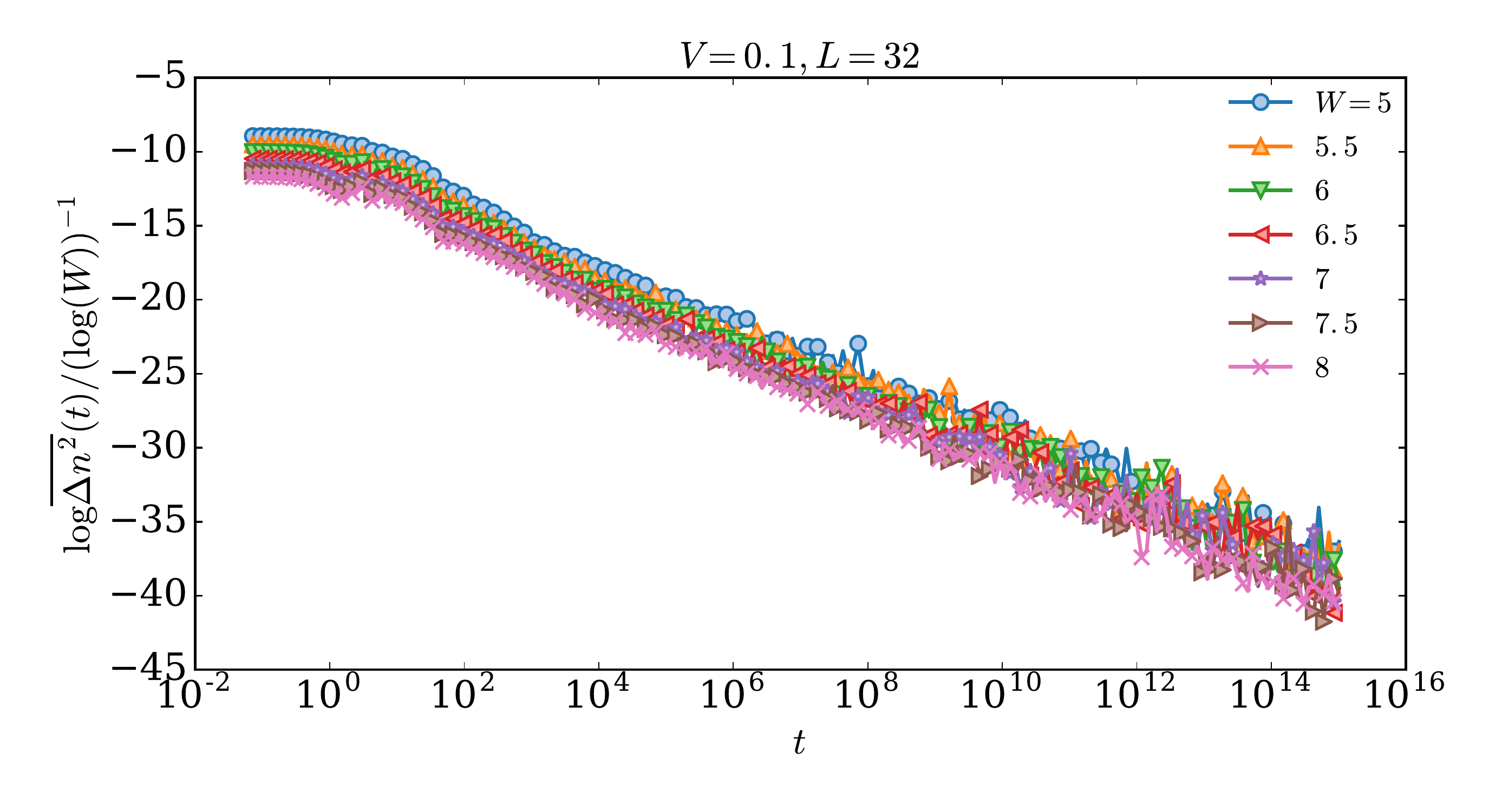}
\caption{$\overline{\Delta n^2}(t)$ calculated using $\hat{H}^{\text{eff}}$ for several $W$'s in the strong disorder limit and fixed $V$ and $L$ with initial state  $|\psi \rangle = \prod_l^L \frac{\hat{\eta}_l + \hat{\eta}_l^\dagger}{\sqrt{2}} | 0 \rangle$.} 
\label{fig:S4}
\end{figure}
It is also important to study the flow of $\mathcal{P}(x)$ with 
system size. Figure~\ref{fig:S6} shows the rescaled probability distribution of both $\mathcal{B}_{l,m}$ and $\mathcal{B}_{l,l,p,s}$. $\mathcal{P}(x)$ for $x=\mathcal{B}_{l,m}$ goes to zero as $1/L$. It is due by the fact that in the strong disorder limit most of $\mathcal{B}_{l,m}$'s ($\sim L^2$)  will have an exponentially small value since $\mathcal{B}_{l,m}$ involve the ``overlap'' between two localized wave functions. Nevertheless, some of them will have a large overlap $(\sim L)$, these terms represent wave functions that are close by localized. As we already discussed most of $\mathcal{B}_{l,l,p,s}$'s (pick of $\mathcal{P}$ shifts to an exponentially small number in $L$ as shown in Fig.~\ref{fig:S6} ($\mathcal{B}_{l,l,p,s}^{\text{typical}} \sim e^{-\alpha L^\delta}$ with $\delta \approx 0.9$). Moreover the amplitude of having a typical value goes also to zero with $L$ ($\mathcal{P}(x\sim x_{\text{max}})\sim 1/L$), while the probability of having large value of $\mathcal{B}_{l,l,p,s}$ goes to zero faster ($\mathcal{P}(x = \log{ |\mathcal{B}_{l,l,p,s}|} \sim 0)  \sim 1/L^2$).      

We also looked for resonances, meaning of the break down of first order perturbation theory and thus the need to go beyond first order in $V$ or to use degenerate perturbation theory. We studied these elements statistically calculating its probability distribution. The probability of having resonances  give us an estimation of error done in taking for eigenstates the ones of the non-interacting case. Up to first order in the interaction strength $V$, the assistant hopping processes ($\mathcal{B}_{l,l,p,s}$) will give a contribution of the form 
\begin{equation}
 \text{first order in $V$} \sim\frac{B_{l,l,p,s}}{(\epsilon_s - \epsilon_p)},
\end{equation}
where $\epsilon_l$'s are the single-particles eigenenergies.
We will say that two energies $\epsilon_p$ and $\epsilon_s$ are in resonance at first order in $V$ if 
\begin{equation}
 \left |\frac{B_{l,l,p,s}}{(\epsilon_s - \epsilon_p)} \right |\ge 1.
\end{equation}
We calculate the cumulative distribution function (CDF) of the distribution function $\mathcal{P}(x)$ of $x= \log{ \left |\frac{B_{l,l,p,s}}{(\epsilon_s - \epsilon_p)} \right |}$
\begin{equation}
 CDF(x) = \int_{-\infty}^x dx \mathcal{P}(x).
\end{equation}
Figure~\ref{fig:S7} shows $1-CDF(x)= \int_{x}^{\infty} dx \mathcal{P}(x) $ for the values of $W$ and $V$ used in the main text. The probability of having resonances ($1-CDF(x\ge0)$) is of the order of $10^{-4}$ (less than $1\%$). This value tells us what is the probability that our approach is inadequate and higher order processes in $V$ should be considered. We repeated this analysis for several value of $W$ (Fig.~\ref{fig:S8}). 

{\it The two-point connected correlation function $\mathcal{C}_x(t)$}---
In this section we provide a more detailed analysis for the dynamics of the two-point correlation function $\mathcal{C}_x(t)$. In particular, we will use this section also to emphasize the necessity to simulate larger system sizes than the ones accessible by exact diagonalization (ED) in order to understand the right behavior of $\mathcal{C}_x(t)$ in the localized phase. 

First, we would like to demonstrate that in a thermal phase $\mathcal{C}_x(t)$ decays to zero with time. Figure~\ref{fig:Sup_C} shows $\mathcal{C}_x(t)$ for a fix distance $x=3$ in the ergodic phase $(W=1)$ of the one-dimensional model studied in main text~\cite{Luitz15}.  
By definition, at $t=0$, $\mathcal{C}_x(t) = 0$ and after a transient $L$-independent propagation, $\mathcal{C}_x(t)$ saturates on long times to a $L$-dependent value, which shows a marked trend towards zero with $L$. Moreover, it is possible to show that $\lim_{L \rightarrow \infty} \mathcal{C}_x(t) \sim e^{-\alpha L}$.

That this eventually yields a vanishing correlation function can be understood simply by the fact that the system is fully ergodic. Indeed, at infinite temperature, as we considered in the main text, the long-time steady state can be described by a random state (e.g. the entanglement entropy after a quantum quench saturates to the Page value $\mathcal{S} \sim L/2\log{2}$), so that we can take a quantum average over a random state which yields
\begin{equation}
 \lim_{t\rightarrow \infty} \mathcal{C}_x(t), \sim e^{-\alpha L}\qquad \alpha >0,
\end{equation}
Consequently, such correlators vanish in an ergodic system at least at infinite temperature.

Instead, Fig.~\ref{fig:Sup_C1} shows $\mathcal{C}_x(t)$ for $x=2$ deep in the MBL phase. In this case the results are obtained using ED to show the importance to be able to simulate larger systems sizes. 

Indeed, at large times one can observe a slight bending down of the long-time value of $\mathcal{C}_x(t)$ for increasing $L$. Nevertheless it is not possible to predict the behavior in the thermodynamic limit due to the limitations in the accessible system sizes with ED. The efficiency of our method allows us now to address much larger system sizes ($L \sim 50$) at arbitrarily large time scales, and thus to give a prediction on the behavior in the thermodynamic limit.

{\it Time fluctuations}---
In this section, we show further data concerning the time fluctuation of local observables (i.e. $\hat{n}_x$), which is defined by 
\begin{equation}
 \Delta n^2(t) = \frac{1}{L} \sum_x \Delta n^2_x(t),
 \label{eq:time_flu}
\end{equation}
where 
\begin{equation}
 \Delta n^2_x(t)  = \left ( \langle \hat{n}_x \rangle (t)  - \langle \hat{n}_x \rangle_{\text{time ave.} }   \right )^2,
\end{equation}
and
\begin{equation}
\langle \hat{n}_x \rangle_{\text{time ave.}}  = \lim_{T\rightarrow \infty} \frac{1}{T} \int_0^T ds \langle \hat{n}_x \rangle (s),
\end{equation}
is the long-time average of $\langle \hat{n}_x \rangle (t)$. In the main text we show that the time fluctuation decay algebraically $\Delta n^2(t) \sim t^{-\alpha}$. Moreover, we claim that $\alpha \propto \xi_{\text{loc}}$~\cite{Ser14}.
Figure~\ref{fig:S3} shows  $\overline{\Delta n^2}(t)$ for several $W$'s in the strong disorder limit. For all inspected disorder strengths $\Delta n^2(t)$ decays algebraically for several order of magnitude, 
the curves have been rescaled to underline that $\alpha \propto \log{W}^{-1}$. 
Indeed in the strong disorder limit $\xi_{\text{loc}}\sim \log{W}^{-1}$, as shown in Fig.~\ref{fig:S3}, where  $\xi_{\text{loc}}$ has been calculated using standard transfer-matrix technique~\cite{0034-4885-56-12-001}. 
Furthermore, in the main text we support the result  $\Delta n^2(t) \sim t^{-c\xi_{\text{loc}}}$ with an analytical argument starting from a different initial state $|\psi \rangle = \prod_l^L \frac{\hat{\eta}_l + \hat{\eta}_l^\dagger}{\sqrt{2}} | 0 \rangle$. 
Figure~\ref{fig:S4} shows  $\overline{\Delta n^2}(t)$ starting from $|\psi \rangle$ for several disorder strengths $W$, giving evidence that  $\Delta n^2(t) \sim t^{-c\xi_{\text{loc}}}$, with $\xi_{\text{loc}}\sim \log{W}^{-1}$.

\end{document}